# Bound-state-in-continuum guided modes in a multilayer electro-optically active photonic integrated circuit platform.

Kyunghun Han[1,2,3], Thomas W. LeBrun[1], Vladimir A. Aksyuk[1,*]

[1]Physical Measurement Laboratory, National Institute of Standards and Technology, Gaithersburg, MD 20899, USA
[2]Department of Chemistry and Biochemistry, University of Maryland, College Park, MD 20742, USA
[3]Theiss Research, La Jolla, CA 92037, USA

Abstract: Bound states in the continuum (BICs) are localized states existing within a continuous spectrum of delocalized waves. Emerging multilayer photonic integrated circuit (PIC) platforms allow implementation of low index 1D guided modes within a high-index 2D slab mode continuum; however, conventional wisdom suggests that this always leads to large radiation losses. Here we demonstrate low-loss BIC guided modes for multiple mode polarizations and spatial orders in single- and multi-ridge low-index waveguides within a two-layer heterogeneously integrated electro-optically active photonic platform. The transverse electric (TE) polarized quasi-BIC guided mode with low, <1.4 dB/cm loss enables a Mach-Zehnder electro-optic amplitude modulator comprising a single straight $Si_3N_4$ ridge waveguide integrated with a continuous $LiNbO_3$ slab layer. The abrupt optical transitions at the edges of the slab function as compact and efficient directional couplers eliminating the need for additional components. The modulator exhibits a low insertion loss of $\approx$ 2.3 dB and a high extinction ratio of 25 dB. The developed general theoretical model may enable innovative BIC-based approaches for important PIC functions, such as agile spectral filtering and switching, and may suggest new photonic architectures for quantum and neural network applications based on controlled interactions between multiple guided and delocalized modes.

In the context of wave dynamics, a local wave state embedded within a continuous spectrum of propagating states, generally, manifests radiative coupling, resulting in energy dissipation or exchange. However, bound states in the continuum, a concept first introduced by von Neumann and Wigner in 1929[1], can circumvent this conventional wisdom when the mode overlap with the radiative states is minimized or even eliminated. This concept has been extensively explored not only in quantum mechanics but also in various other wave phenomena, including acoustic wave[2], water wave[3], and elastic wave in solid[4]. In the case of optical waves, control over geometry at the wavelength scale allows precise and local engineering of the desired coupling strength as a function of wavelength, mode order and polarization. Recent advancements in integrated photonics nanofabrication enabled the exploration of a diverse range of BICs in various nanophotonic structures, including 2D planar structures[5–9], gratings[10–12], and waveguides [13–17]. In the absence of other dissipation mechanisms, reduced radiative coupling results in sharp spectral features and long-lived bound states. The unique behavior of BICs has been harnessed in various applications, including lasing[5,18,19], and sensing[20–22]. Beyond photonics, this can also describe a quantum mechanical system interacting with an external open environment[23], a system that has many interests such as non-Markovian transition[24,25] and non-Hermitian many body systems[26].

The BICs in the form of 1D waveguide modes first theoretically predicted in 1979[27] and experimentally demonstrated in a radiofrequency waveguide[28,29] and photonic waveguide[30]. A BIC waveguide mode can be manifest by a conventional ridge waveguide in conjunction with a nearby 2D slab waveguide when the effective index of a ridge mode is below that of the slab modes. While the guided mode generally couples with, and dissipates into, the set of slab modes that satisfy the phase matching condition, this dissipation can be suppressed under certain waveguide width conditions. However, presently the photonic BIC guided mode demonstrations have been limited to orthogonal polarizations between the guided mode and the slab modes within a polymer ridge platform[14–16,31–33] or a rib waveguide[34]. There has not been any demonstration of the BIC waveguiding in the same polarizations between the guided mode and the radiating slab modes (e.g., quasi-TE). Lacking a systematic and general description, the applicability of BIC waveguiding to different modal polarizations and spatial geometries in PIC platforms using silicon, silicon nitride as well as different electro-optic and nonlinear materials and their combinations, has not been fully explored.

Here, we experimentally demonstrate photonic BICs in single and multiple waveguide configurations and develop a general theoretical description. We establish low loss BIC waveguiding for both quasi-TE and quasi-TM polarizations and multiple spatial modes. For the practical usage of BICs, our experimental platform comprises a thin-film lithium niobate (TFLN) layer heterogeneously integrated on top of a silicon nitride PIC. We realize a Mach-Zehnder electro-optic amplitude modulator with the quasi-BIC waveguide mode serving as a low-loss reference channel effectively decoupled from the lithium niobate (LiNbO$_3$) layer and insensitive to electro-optic modulation. The simple design with the single straight waveguide and the slab layer is further enabled by using the inherent abrupt modal transitions at the TFLN slab edges, which act as directional couplers for interferometry. The modulator exhibits ≈ 25 dB extinction ratio and an electro-optic half-wave voltage-length product of ≈ 10.8 V·cm, an improvement factor of ≈ 7.4. [15,35,36]. The improvement is due to the quasi-transverse electric (TE) polarization for both the BIC mode and the fundamental mode, leveraging the highest electro-optic coefficient in the x-cut lithium niobate (r$_{33}$).

## Bound-states-in-continuum guided mode

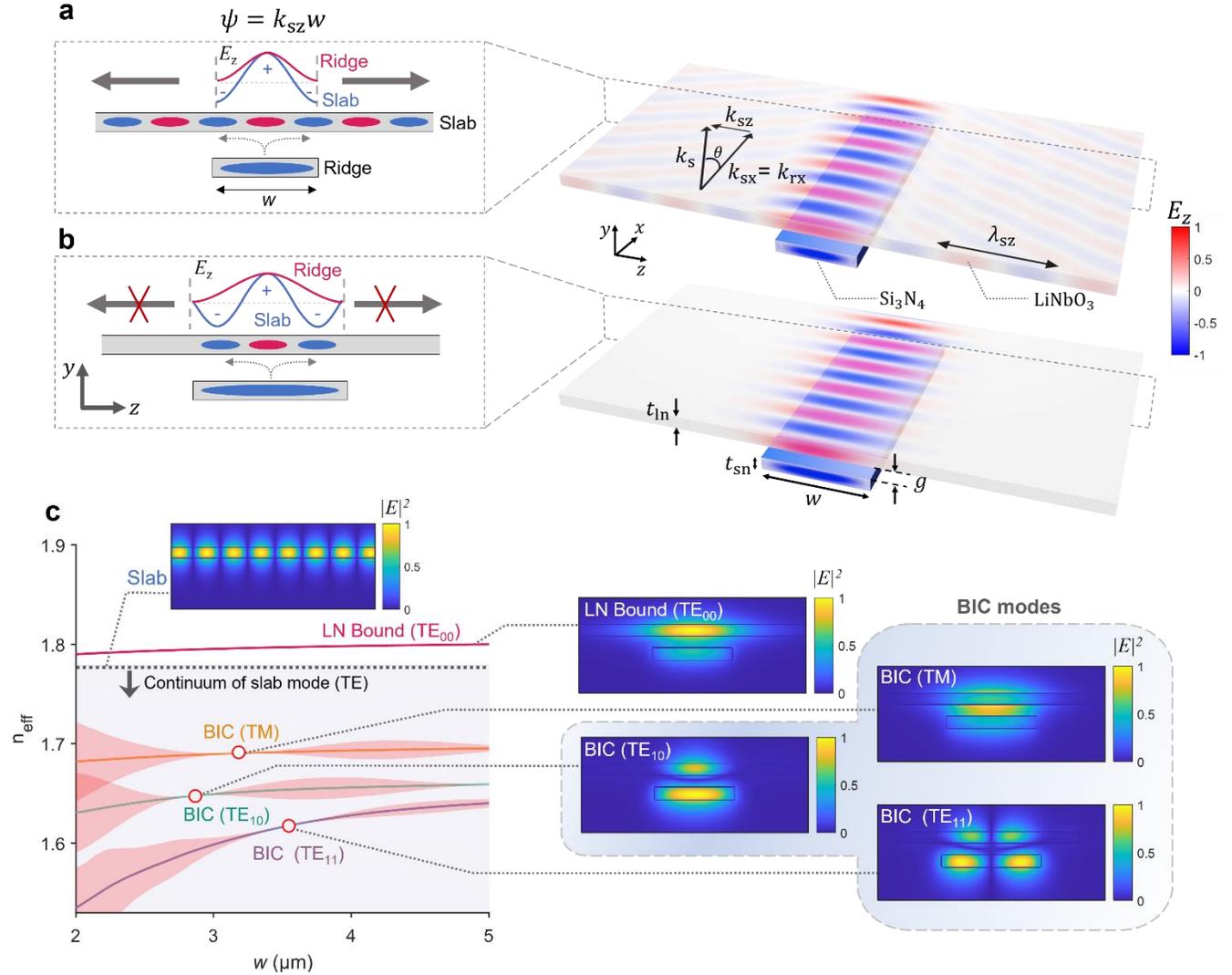

**Fig. 1 Bound states in continuum modal properties**. **a,** Schematic illustrations of a single ridge waveguide mode (Si$_3$N$_4$) interacting with a nearby slab waveguide mode (LiNbO$_3$), along with corresponding 3D FDTD simulation results showing the electric field $E_z$ distributions at a leaky resonant condition. The slab waveguide mode, which is phase-matched with the hybridized ridge waveguide mode ($k_{rx}=k_{sx}$), efficiently couples and dissipates the energy. **b,** Lateral wave dissipations are blocked when a lateral phase shift $\psi=k_{sz}w$ of the slab mode satisfies the BIC condition, $\psi \approx (2m+1)\pi$. This indicates a net zero overlap between the ridge and the slab modes. **c,** Simulated effective refractive index ($n_{\text{eff}}$) and corresponding electric field $|E|^2$ of the LN bound mode (red), TM BIC mode (orange), TE$_{10}$ BIC mode (green), and TE$_{11}$ BIC mode (purple) at $g$=300 nm, $t_{sn}$=350 nm, $t_{ln}$=300 nm, at the vacuum wavelength $\lambda_0$=1550 nm. The shaded region indicates the existence of the continuous spectrum of the TE slab modes, where any bound mode within this region could possibly become the leaky resonant modes. The width of the red shaded regions along the refractive index curves indicates the imaginary part of the effective indices, scaled independently for each mode by setting their maximum values at $w$= 2 µm to 0.08.

BICs in the heterogeneously integrated thin-film lithium niobate platform can exist in a single, a double or, generally, a multiple-ridge waveguide. We first consider a single rectangular silicon nitride (Si$_3$N$_4$) buried channel waveguide in proximity to the continuous thin-film lithium niobate (LN) slab (Fig. 1), establishing a hybridized ridge waveguide mode. For the purposes of this manuscript, all polarizations should be understood as 'quasi-' even if the prefix is not explicitly mentioned. Hybridization of the slab modes with the buried channel waveguide modes leads to the emergence of a lossless guided LN bound $TE_{00}$ mode within the slab, while the other hybridized modes overlap with the slab mode continuum (Fig. 1c), since their various wavevectors (e.g., $k_r$) are smaller than the slab mode wavevector $k_s$. The ridge waveguide mode becomes lossy through radiation into the slab modes, with the strongest coupling to the modes propagating at the angle $\theta$ to the waveguide satisfying the phase matching condition $k_{rx} = k_{sx} = k_s \cos(\theta)$ (Fig. 1a). $k_{rx}$ and $k_{sx}$ represent wavevectors along the x-axis for the ridge and the slab waveguide modes, respectively.

In the cross-section normal (y-z plane) to the propagation direction (x-axis), the slab mode is a laterally oscillating wave with a wavelength $\lambda_{sz} = 2\pi/\sqrt{k_s^2 - k_{sx}^2}$ along the z-axis. Meanwhile, the ridge modes have a lateral structure set by the waveguide width for each polarization and transverse modal order. Therefore, the mode overlap integral between the waveguide and the slab modes is an oscillating function of the waveguide width containing multiple zero-crossings. At these zero-crossings the hybrid ridge mode is a BIC, uncoupled from the radiating slab continuum and the lateral dissipation from it is prohibited (Fig. 1b). As illustrated in Figure 1c, for the hybrid waveguide modes of various polarizations and spatial orders become BIC at specific waveguide widths.

For the transverse magnetic (TM) hybrid mode, the mode overlap with the TE slab continuum is much smaller for most waveguide widths. However, the mode overlap is not fully suppressed by polarization orthogonality alone because of a broken vertical index symmetry[14,32,34,37] (See section B of Supplementary material).

The BIC condition described here fundamentally arises from the oscillating structure of the phase-matching modes in the slab continuum, and therefore it is not limited to single ridge waveguides, but occurs with any guided modes that can be laterally scaled with the waveguide geometric structure. This will be further illustrated below in a specific case of two ridge waveguides coupled through the slab.

The interaction between a bound mode and radiating slab modes can be theoretically modeled using coupled mode theory (CMT)[23,38,39]. If the slab waveguide width ($s$) in the z-axis is sufficiently large that the coupled energy is dissipated without reflection, the interaction can be viewed as one oscillator (bound mode) and adjacent multiple radiating modes (slab modes) that carry coupled energies. The interaction along the propagation direction (x-axis) can be described by the coupled mode equations (See section B of Supplementary material). The power propagation exponential loss length $L$ can be expressed as

$$\frac{1}{L} = \gamma_0 n_{eff} \lambda_{sz}^5 \frac{w_{eff} \cos^2\left(\frac{\pi}{\lambda_{sz}} w_{eff}\right)}{\left(4 w_{eff}^2 - \lambda_{sz}^2\right)^2} \qquad (1)$$

where $\gamma_0$ is a scaling factor that accounts for the y-dependency in the integrals for the coupling coefficients. $w_{eff} = \lambda_{rz}/2$ and $\lambda_{rz} = 2\pi/\sqrt{k_f^2 - k_{rx}^2}$ are a function of the waveguide width, considering the waveguide dispersion. The z-dependency of the ridge waveguide electric field can be modeled as $\cos(2\pi/\lambda_{sz} z)$ where $|z| < \lambda_{sz}/4$. $n_{eff}$ is an effective refractive index of the hybridized ridge waveguide mode. This equation indicates that the BIC condition is satisfied at multiple waveguide widths where the cosine term in the numerator is zero. These BICs originate from the mode overlap between two waveguide modes becoming zero at $w_{eff} = (m + 1/2)\lambda_{sz}$, with a positive integer $m$. There is no BIC condition for $m=0$ due to the denominator term.

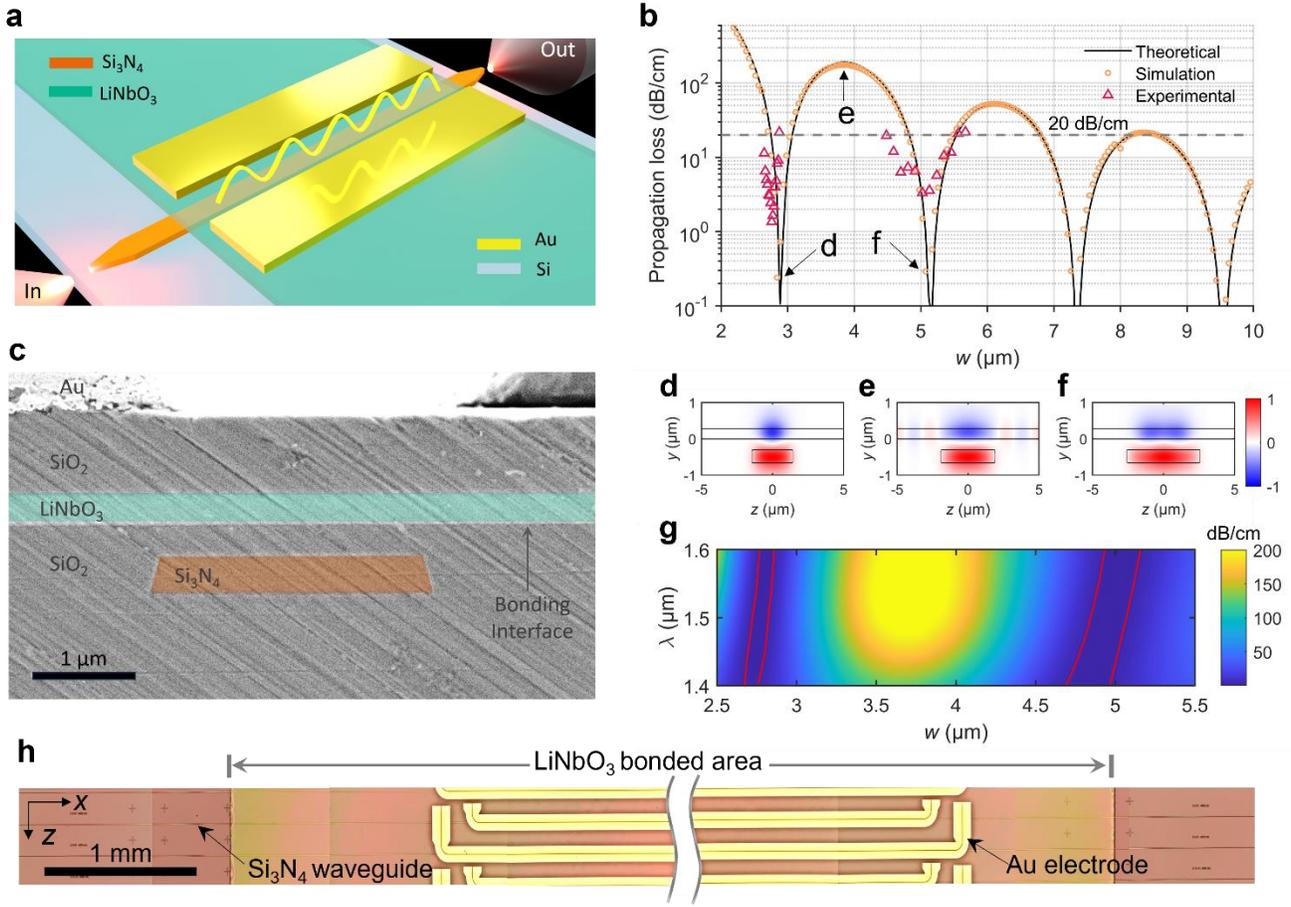

**Fig. 2 Demonstration of TE$_{10}$ BIC mode in heterogeneously integrated lithium niobate hybrid waveguide**. **a,** Illustration of the fabricated waveguide. The input light, injected into the silicon nitride waveguide, meets the lithium niobate bonded area where the refractive index is modulated by the external electric field. **b,** Propagation loss for different waveguide widths. Black solid line is coupled mode theory, orange circles are finite element eigenmode numerical analysis, and red triangles are experimental results. There are multiple quasi-BIC conditions with a period of $\lambda_{sz}$ (≈2.2 µm). The 20 dB/cm dashed line represents the highest experimentally-measurable propagation loss, limited by the ability to separate the low-loss fundamental mode. Uncertainties of individual measurements are estimated to be smaller than the symbol size. **c,** False-colored scanning electron microscopy image of polished cross-section for the fabricated waveguide, showing a defect-free bonding interface. **d-f,** Numerically calculated electric field ($E_z$) profiles for the first BIC condition, the leaky resonant condition, and the second BIC condition. **g,** Numerically calculated propagation loss as a function of a waveguide width ($w$) and the optical vacuum wavelength ($\lambda$). The condition where the propagation loss is at 3 dB/cm is colored in red. **h,** Stitched optical microscope image of the fabricated device where lithium niobate is bonded in the middle of the chip (nominal length of 17 mm) while the rest of area is free from lithium niobate. The gold electrode has a nominal thickness of 900 nm and a nominal gap of 2.85 µm.

Heterogeneous integration of TFLN with PIC platforms[33–37] is a cost effective and flexible strategy whereby TFLN is added only to the areas utilized for active modulation, leaving the remainder available for passive PIC components and, potentially, hybrid integration of distinct active layers for lasers and detectors. In our hybrid waveguide (Fig. 2a), we used an x-cut TFLN with a horizontal configuration of Au electrode on top, to exploit the highest electro-optic coefficient ($r_{33}$) (Fig. 2h). The measured loss in the inversely tapered fiber-to-chip edge couplers optimized for lensed fiber is (1.55 ± 0.51) dB per facet (all uncertainties are one standard deviation statistical uncertainties unless noted otherwise). This is a strong advantage of heterogenous integration over the conventional etched rib waveguides with TFLN, which present the fiber-to-chip coupling loss from 5 dB to 10 dB[40]. A defect-free TFLN layer bonding (Fig. 2c) is achieved using nominally 3 nm of atomic layer deposited (ALD) $Al_2O_3$ as an intermediate layer (see section A of Supplementary Material).

The experimentally measured TE$_{10}$ BIC mode propagation losses (Fig. 2b) are in agreement with the finite element eigenmode analysis simulation and the theoretical prediction from Eq.(1), clearly showing the minima for the waveguide widths $w_{\text{eff}} = (m + 1/2)\lambda_{sz}$. The method to characterize the experimental propagation loss is described in the section C of the Supplementary Material. The functional form of the propagation loss is $\propto \cos^2(\pi w_{\text{eff}}/\lambda_{sz})/w_{\text{eff}}^3$. While low propagation losses have been previously shown[15] for TM polarized modes within the TE polarized slab continuum, here we show that propagation losses <1.5 dB/cm can be experimentally achieved for TE polarization, without relying on polarization orthogonality for the low mode overlap. Moreover, our modeling (Fig. 2g) indicates that the propagation loss of a TE quasi-BIC mode at wavelengths around 1550 nm can be less than 3 dB/cm over a wavelength bandwidth of up to ≈100 nm (Fig. 2g) and waveguide width variation of up to ≈100 nm. Overall, the quasi-BIC condition for a single waveguide exhibits robustness against fabrication and wavelength variations while maintaining the low loss propagation without requiring polarization orthogonality.

## BIC-based Mach-Zehnder amplitude modulator

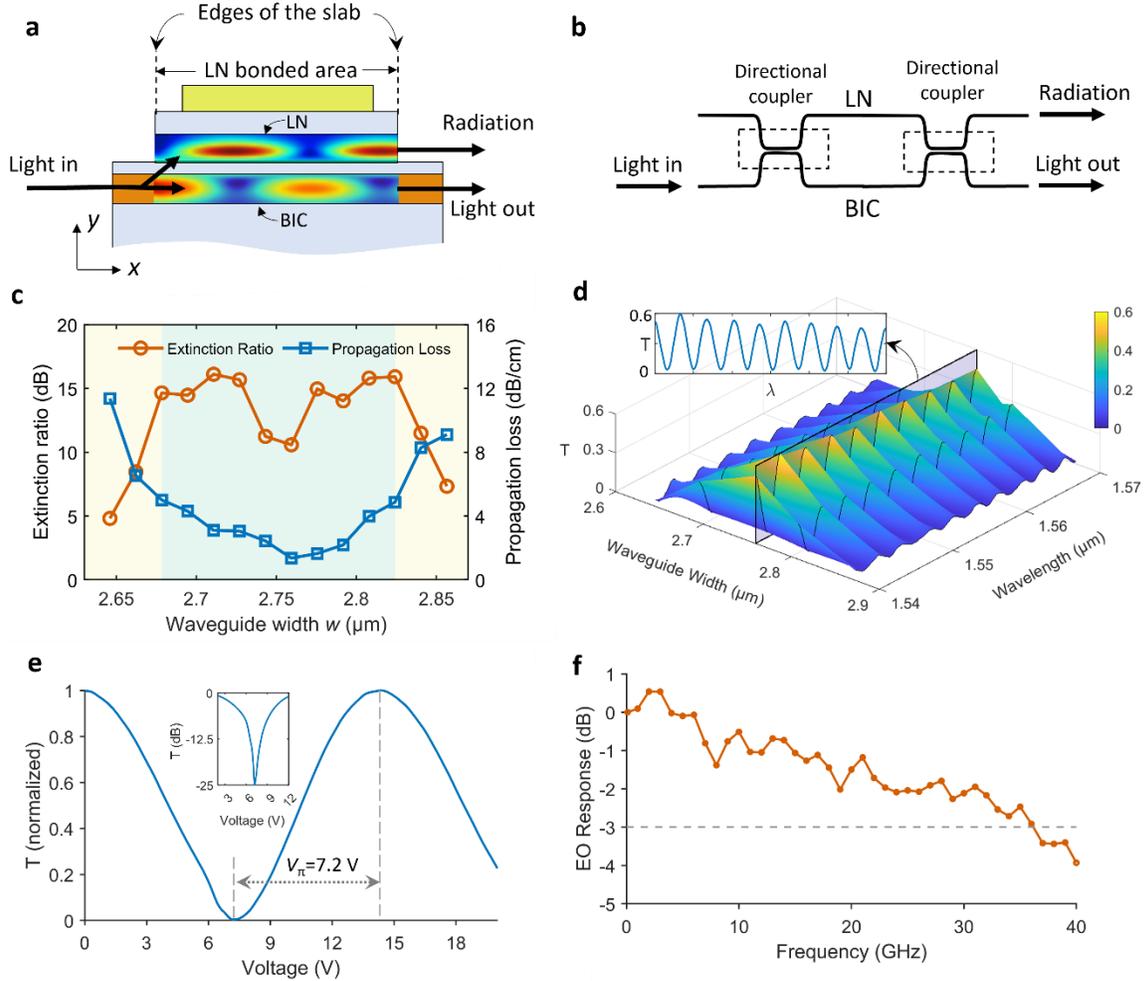

**Fig. 3 Mach-Zehnder modulator based on quasi-BIC a,** Schematic of the BIC-MZI modulator. At the edge of the bonded lithium niobate slab, the input light is coupled into two waveguide mode channels due to the abrupt transition in the modal bases, inherently serves as the directional couplers. The electric field profile shows a multi-mode beating pattern with an oscillating power transfer between two waveguides. **b,** Equivalent BIC-MZI model with two asymmetric channels with different effective refractive indices but the same physical length. LN stands for the lithium niobate confined bound mode channel. BIC stands for the $TE_{10}$ BIC waveguide mode channel. The output lensed fiber collects the light from the light out port. The two directional couplers represent the two bonding edges in the LN slab. **c,** Measured average values of the extinction ratios over the transmission spectra with various waveguide widths (orange circles with the left y-axis), and corresponding characterized propagation losses (blue circles with the right y-axis). The boundaries of the two shaded areas indicate the powers from the two channels are balanced at the light out port, resulting in a high extinction ratio. **d,** Measured the total transmission, including the fiber-to-chip coupling efficiency, of the BIC-MZI as a function of the waveguide width and the wavelength. The inset shows the transmission spectrum of the device with the highest transmission. **e,** Measured optical transmission by electro-optic modulation of the BIC-MZI modulator at the wavelength giving the highest extinction ratio. A sawtooth waveform was applied as a driving radio frequency (RF) signal to evaluate a half-wave voltage. The inset shows the log scale transmission indicating 25 dB extinction ratio. **f,** Measured electro-optic response with various modulation frequencies, characterized with the RF spectrum analyzer. The 3 dB bandwidth was measured at ≈36 GHz.

To demonstrate the potential of BICs, the Mach-Zehnder interferometer (MZI) is realized with a simple straight silicon nitride waveguide with a bonded lithium niobate slab (Fig. 3a). In the absence of an adiabatic transition, the waveguide mode experiences strong modal perturbation at the first edge of the lithium niobate slab, leading to coupling into a new basis set of waveguide eigenmodes. The powers coupled into these new eigenmodes, shown in Fig. 1c, are determined by their mode overlaps with the $TE_{00}$ mode of the input $Si_3N_4$ buried channel waveguide. The simulation shows that most power couples into either the fundamental TE$_{00}$ LN bound mode (refer to as the LN channel) (32 %) or quasi-BIC TE$_{10}$ waveguide mode (refer to as BIC channel) (62 %), while the remaining power becomes a scattering loss. We intentionally biased the split ratio for a higher power in the BIC channel due to its higher propagation loss compared to that of the LN channel. After propagation the modes encounter the second bonding edge, where they either couple back into the silicon nitride ridge waveguide (output port) or scatter out into free-space (radiation port). Thus the abrupt transitions serve as simple and compact directional couplers, circumventing conventional design challenges in heterogenous integration of the TFLN, requiring additional structures for an adiabatic transition such as the tapered structure[40] or the varying waveguide width with a higher effective index than the lithium niobate[41–43] to avoid mode mismatch loss at the edges.

The described BIC-based MZI (BIC-MZI) is functionally analogous to an MZI with asymmetric channels (Fig. 3b): the fundamental TE$_{00}$ mode for the upper channel and the quasi-BIC waveguide mode for the lower channel. These waveguide eigenmodes originate from the hybridization of the coupled buried channel and slab TE modes, resulting in an apparent electric field beating pattern as the modes propagate (Fig. 3a). Given the wavelength-dependent phase shift due to the asymmetric channels, the transmission spectrum of the BIC-MZI modulator has an oscillation (Fig. 3d). The highest transmission at an optimized waveguide width and optical wavelength was measured at ≈0.59, which is equivalent to the intrinsic modulator insertion loss of ≈2.3 dB. This fully includes losses at the Si$_3$N$_4$ buried channel waveguide to TFLN/Si$_3$N$_4$ hybrid waveguide transitions, and only excludes the fiber-to-chip coupling loss.

It is difficult to separately measure each mode's propagation loss via conventional cut-back method. We extracted the propagation losses of the BIC modes for various waveguide widths from the measured transmission (Fig. 3c), using an analytic model of the MZI (see section C of Supplementary material). The lowest measured propagation loss was (1.36 ± 0.28) dB/cm. The minimum propagation loss in Fig. 3c does not necessarily correspond to the maximum extinction ratio because of the unequal propagation loss in the two channels and the unequal split ratio of the directional couplers. In Fig.3c, the blue-shaded regions denote instances where BIC channel power is estimated to exceed the LN channel at the light out port, while the orange-shaded regions indicate weaker BIC channel power. The boundary between them represents balanced powers from both channels, leading to a higher extinction ratio.

Electro-optic amplitude modulation was achieved by applying a sawtooth waveform to the Au electrode (Fig. 3e). Because the electric field density of the BIC optical mode is mostly confined in the silicon nitride waveguide and decoupled from the lithium niobate, the mode's effective refractive index change due to the applied electric field is smaller compared to that of LN mode (see section C of Supplementary material). A half-wave voltage was measured at ≈7.2 V in a nominally 15 mm long modulation region, equivalent to ≈10.8 V·cm with a nominal electrode gap of 2.85 μm. Although this half-wave voltage is higher than that of state of the art TFLN modulators, it is almost 7.4 times lower compared to the previously reported BIC-based modulator with the TM polarized waveguide mode [15,35,36]. This improvement is due to the highest electro-optic efficiency obtained when the lithium niobate crystallographic axis, applied electric field and optical polarization are oriented along the same (z) axis in our design. The extinction ratio was maximized by selecting an optimal wavelength and measured at ≈25 dB. The extinction ratio generally depends on achieving the precise power balance at the modulator output and can also be limited by challenges in achieving pure linear polarization[44] and the existence of unwanted modes. The electro-optic response of the BIC-MZI was characterized over a modulation frequency range from 0.1 GHz to 40 GHz by a radio frequency (RF) spectrum analyzer (See section C of Supplementary material). The 3 dB bandwidth of the BIC-MZI is measured to be ≈36 GHz (Fig. 3f). This bandwidth could be further improved by employing a capacitively-loaded coplanar waveguide design[45].

# Double waveguide BIC

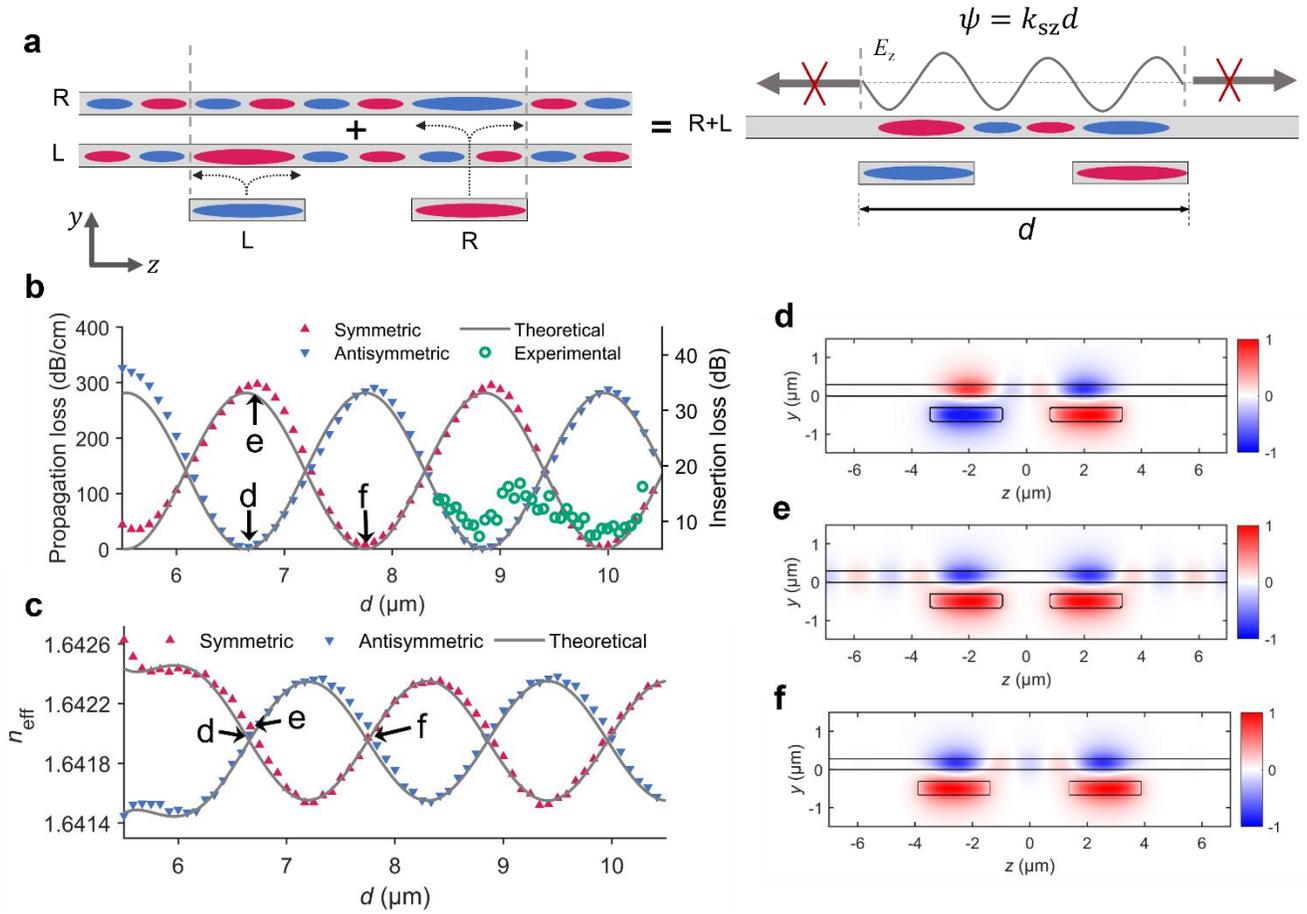

**Fig. 4 Double waveguide BIC modal properties**. **a,** Schematic illustrations of two ridge waveguides ($Si_3N_4$) with 180° out-of-phase condition, interacting with a nearby slab waveguide ($LiNbO_3$). The dissipating waves from the left-side ridge waveguide (L) and the right-side ridge waveguide (R) interfere, resulting in destructive interferences outside the edges of the two ridge waveguides (grey dashed lines). The complete destructive interference, a BIC condition, occurs when the phase shift along z-axis, $\psi = k_{sz}d$, satisfies $\psi = (2m)\pi$ for an anti-symmetric mode and $\psi = (2m+1)\pi$ for a symmetric mode. **b-c,** Propagation losses and **e**ffective refractive indices and of the anti-symmetric and symmetric modes calculated by the coupled mode theory (black solids), eigenmode analysis (red and blue triangles), and experimental result (green circles) for $w$=2.5 μm. From the perspective of supermodes, BICs are results of two oscillators interactions coupled through radiating channels showing the oscillating patterns in the effective indices and the propagation losses. The experimental result represents the on-chip insertion loss without the fiber-to-chip losses with y-axis on the right. Uncertainties of individual measurements are estimated to be smaller than the symbol size. The rest of the data and curves are plotted with the y-axis on the left. **d-f,** z-component of the electric field ($E_z$) profile for the anti-symmetric BIC condition, symmetric leaky resonant condition, and symmetric BIC condition

    In the case of a double, or multiple, waveguide configuration, BICs may be achieved for specific supermodes by only varying the waveguide spacings between two adjacent waveguides (*d*). It is no longer necessary that the individual waveguide widths (*w*) satisfy the BIC condition when multiple waveguides are coupled via the slab. Instead, for weak slab coupling these waveguides can be simply understood as individual oscillators, each leaking into a common dissipating channel[4], the lithium niobate layer (Fig.4a). The radiating waves from the left-side (L) and the right-side waveguide (R) may destructively interfere at the outside edges of the two ridge waveguides, leading to the double waveguide BICs.

    Quantitatively, the behavior of the two oscillators can be modeled by a coupled mode theory. The coupled mode equations can be written as

$$\frac{1}{k_0}\frac{\partial A}{\partial x} = -i(n_{\text{eff}} - iG)A - i(K - iGe^{i\psi})B \tag{2}$$

$$\frac{1}{k_0}\frac{\partial B}{\partial x} = -i(n_{\text{eff}} - iG)B - i(K - iGe^{i\psi})A \tag{3}$$

where $n_{\text{eff}}$ is the effective index of the single waveguide mode, $K$ is a coupling coefficient of the evanescent wave coupling. $G$ is a radiation rate of the resonances into the coupling channel, which is the same value with the coupling coefficient of the single waveguide BIC mode. $\psi$ is a phase distance between the two resonances along the radiation channel, defined by how far apart they are. The effective index eigenvalues for the supermode eigenvectors can be written as $n_{\pm} = n_{\text{eff}} \pm K - iG(1 \pm e^{i\psi})$, where '+' sign is for the symmetric supermode and '-' sign is for the anti-symmetric supermode. The BICs are satisfied when the phase shift ($\psi$) along the channel in z-direction between two waveguides, defined as $\psi = k_{sz}d$, becomes $(2m)\pi$ for 180° out-of-phase oscillators (anti-symmetric) or $(2m+1)\pi$ for in-phase oscillators (symmetric), where $m$ is the positive integer. These conditions are Fabry-Pérot type BICs, showing good agreement with the simulated result and the experimental result (Fig. 4b). The propagation losses of the two waveguide modes, the symmetric and the anti-symmetric supermode, are proportional to $\cos^2(k_{sz}d/2)$ and $\sin^2(k_{sz}d/2)$, respectively. Due to the complicated interferometer model, we used the maximum transmission in each spectrum of the devices with various widths and extracted the on-chip insertion loss excluding the fiber-to-coupling loss to show the existence of the BIC. The waveguide width giving quasi-BIC conditions are well-matched, validating the model for the double waveguide quasi-BIC modes.

The effective indices oscillate as the waveguide spacing increases (Fig. 4c). There are two distinct regimes for coupling to be considered. In the initial regime, occurring when the waveguides are in proximity ($d<6$ μm), coupling is primarily driven by evanescent wave coupling. As the spacing extends beyond this point, the coupling through the channel overtakes. This regime shows clear oscillations due to the interaction of the two waveguide modes. Consequently, the BIC conditions are located where two effective refractive indices of two waveguide supermodes become equal.

The double waveguide configuration not only provides a framework realizing the multi-waveguide BIC condition but also offers potential for many applications. The numerical simulation suggests the existence of an exceptional point in the double waveguide configuration (see section B of Supplementary material). Furthermore, this can be a great framework as a photonic simulation for many-body quantum system with an open environment where dissipation is a key component[26].

## Discussion and conclusion

In summary, we theoretically and experimentally demonstrated low-loss BIC guided mode within the heterogeneously integrated electro-optically active photonic platform. For demonstrations, the TE polarized quasi-BIC guided mode with ≈1.4 dB/cm loss served as a reference channel in a Mach-Zehnder amplitude modulator comprising a single $Si_3N_4$ ridge waveguide integrated with a continuous $LiNbO_3$ slab layer. The modulator simplifies device architecture by utilizing the abrupt transitions from heterogeneous integration, which acts as directional coupler, eliminating the need for additional components to avoid the mode mismatch loss. The modulator with the optimized design and wavelength resulted in the insertion loss of ≈2.3 dB and an extinction ratio of ≈25 dB. Our findings illuminate the unexplored potential of BICs in PICs, paving the way for significant advancements in fields such as lithium niobate-based quantum photonics[46–48], and photonic neural network[49–51]. This work thus serves as a steppingstone towards the integration and exploitation of BICs in diverse photonic applications, enhancing our ability to manipulate and control light at the nanoscale.